\documentclass[showpacs,english,twocolumn,amsmath,amssymb,aps,prl, superscriptaddress,floatfix, sort&compress, altaffillsymbol]{revtex4-1}
\usepackage[T1]{fontenc}
\usepackage[latin1]{inputenc}
\usepackage{graphicx}
\usepackage{amssymb}
\usepackage{verbatim}
\usepackage{epic,eepic}
\usepackage{babel}
\usepackage{lmodern}
\usepackage{color}

\begin{document}

\title{Signals of Bose Einstein condensation and Fermi quenching in the decay of hot nuclear systems}

\author{P.~Marini}\email{marini@cenbg.in2p3.fr}
    \altaffiliation{Present address: Centre d'Etudes Nucl\'eaires de Bordeaux Gradignan, Chemin du Solarium Le Haut Vigneau, 33175 GRADIGNAN, France}
   \affiliation{Grand Acc\'el\'erateur National d'Ions Lourds, Bd Henri Becquerel BP 55027 - 14076 CAEN, France}
   
\author{H.~Zheng }
\affiliation{Cyclotron Institute, Texas A$\&$M University, College Station, TX-77843, USA}
\affiliation{Physics Department, Texas A$\&$M University, College Station, TX 77843, USA}

\author{M. Boisjoli}
\affiliation{Laboratoire de Physique Nucl\'eaire, Universit\'e Laval, Qu\'ebec, Canad G1K 7P4}
\affiliation{Grand Acc\'el\'erateur National d'Ions Lourds, Bd Henri Becquerel BP 55027 - 14076 CAEN, France}

\author{G. Verde}
      \affiliation{Institut de Physique Nucl\'eaire, IN2P3-CNRS, Universit\'e Paris-Sud 11, F-91406 Orsay Cedex, France}
 \affiliation{INFN- Sezione di Catania, via Santa Sofia, 62, 95123 Catania, Italy}

\author{A. Chbihi}
    \affiliation{Grand Acc\'el\'erateur National d'Ions Lourds, Bd Henri Becquerel BP 55027 - 14076 CAEN, France}

\author{G. Ademard}
    \affiliation{Institut de Physique Nucl\'eaire, IN2P3-CNRS, Universit\'e Paris-Sud 11, F-91406 Orsay Cedex, France}

\author{L. Auger}
\affiliation{Laboratoire de Physique Corpusculaire, ENSICAEN, Universit\'e de Caen Basse Normandie, CNRS/IN2P3, F-14050 Caen Cedex, France}

\author{C.~Bhattacharya}
    \affiliation{Variable energy cyclotron center, Kolkata, India}

\author{B. Borderie}
    \affiliation{Institut de Physique Nucl\'eaire, IN2P3-CNRS, Universit\'e Paris-Sud 11, F-91406 Orsay Cedex, France}

\author{R. Bougault}
\affiliation{Laboratoire de Physique Corpusculaire, ENSICAEN, Universit\'e de Caen Basse Normandie, CNRS/IN2P3, F-14050 Caen Cedex, France}

\author{J. Frankland}
    \affiliation{Grand Acc\'el\'erateur National d'Ions Lourds, Bd Henri Becquerel BP 55027 - 14076 CAEN, France}
    
\author{E. Galichet}
    \affiliation{Institut de Physique Nucl\'eaire, IN2P3-CNRS, Universit\'e Paris-Sud 11, F-91406 Orsay Cedex, France}

\author{D. Gruyer}
    \affiliation{Grand Acc\'el\'erateur National d'Ions Lourds, Bd Henri Becquerel BP 55027 - 14076 CAEN, France}

\author{S. Kundu}
    \affiliation{Variable energy cyclotron center, Kolkata, India}

\author{M.~La~Commara}
\affiliation{Dipartimento di Fisica, Università di Napoli "Federico II", Napoli (Italy)}
\affiliation{INFN - Sezioni di Napoli, Napoli (Italy)}

\author{I. Lombardo}
\affiliation{Dipartimento di Fisica, Universit\`a di Napoli "Federico II", Napoli (Italy)}
\affiliation{INFN - Sezioni di Napoli, Napoli (Italy)}

\author{O. Lopez}
\affiliation{Laboratoire de Physique Corpusculaire, ENSICAEN, Universit\'e de Caen Basse Normandie, CNRS/IN2P3, F-14050 Caen Cedex, France}

\author{G. Mukherjee}
    \affiliation{Variable energy cyclotron center, Kolkata, India}

\author{P. Napolitani}
    \affiliation{Institut de Physique Nucl\'eaire, IN2P3-CNRS, Universit\'e Paris-Sud 11, F-91406 Orsay Cedex, France}

\author{M. Parlog}
    \affiliation{LPC Caen, ENSICAEN, Universit\'e de Caen, CNRS-IN2P3, Caen, France}

\author{M.~F.~Rivet}
     \affiliation{Institut de Physique Nucl\'eaire, IN2P3-CNRS, Universit\'e Paris-Sud 11, F-91406 Orsay Cedex, France}

\author{E. Rosato}
\affiliation{Dipartimento di Fisica, Università di Napoli "Federico II", Napoli (Italy)}
\affiliation{INFN - Sezioni di Napoli, Napoli (Italy)}

\author{R. Roy}
\affiliation{Laboratoire de Physique Nucl\'eaire, Universit\'e Laval, Qu\'ebec, Canada G1K 7P4}

\author{G. Spadaccini}
\affiliation{Dipartimento di Fisica, Università di Napoli "Federico II", Napoli (Italy)}
\affiliation{INFN - Sezioni di Napoli, Napoli (Italy)}

\author{M. Vigilante}
\affiliation{Dipartimento di Fisica, Università di Napoli "Federico II", Napoli (Italy)}
\affiliation{INFN - Sezioni di Napoli, Napoli (Italy)}

\author{P. C. Wigg}
\affiliation{University of Liverpool
School of Physical Sciences, Physics Department, Oliver Lodge Laboratory, Oxford Street, Liverpool L69 7ZE, UK}

\author{A.~Bonasera}
\affiliation{Cyclotron Institute, Texas A$\&$M University, College Station, TX-77843, USA}
\affiliation{Laboratori Nazionali del Sud, INFN, via Santa Sofia, 62, 95123 Catania, Italy}

\collaboration{INDRA Collaboration}

\date{\today}

\begin{abstract}
We report experimental signals of Bose-Einstein condensation in the decay of hot Ca projectile-like sources produced in mid-peripheral collisions at sub-Fermi energies. The experimental setup, constituted by the coupling of the INDRA 4$\pi$ detector array to the forward angle VAMOS magnetic spectrometer, allowed us to reconstruct the mass, charge and excitation energy of the decaying hot projectile-like sources. Furthermore, by means of quantum fluctuation analysis techniques, temperatures and mean volumes per particle {\it as seen by} bosons and fermions separately are correlated to the excitation energy of the reconstructed system. The obtained results are consistent with the production of dilute mixed (bosons/fermions) systems, where bosons experience a smaller volume as compared to the surrounding fermionic gas. Our findings recall similar phenomena observed in the study of boson condensates in atomic traps.  
\end{abstract}

 
\maketitle

The study of quantum systems composed of mixtures of bosons and fermions stimulates significant theoretical and experimental efforts in different fields of physics. For instance investigations on mixtures of bosonic and fermionic atomic quantum systems, with the outstanding example of $^{3}$He-$^{4}$He fluids, have led to the observation of Bose Einstein condensation (BEC) and Fermi quenching (FQ) \cite{ebner1971,schreck2001}. Atomic nuclei are commonly described  as systems  of strongly interacting fermions, namely neutrons and protons. However, in experiments we observe the existence of phenomena that can be explained 
by considering nuclei as systems composed of bosonic clusters, the most common being $\alpha$ ($^4$He) particles. Among these phenomena we mention the predominance of $\alpha$ radioactivity in heavy nuclei, 
the presence of preformed $\alpha$ in the ground  state of nuclei \cite{scarpaci2010} and the cluster structure of light N=Z nuclei \cite{freer2007, KanadaPRC89}. In the $60$'s the microscopic $\alpha$ cluster model \cite{bertsch71,fujiwara80} succedeed in describing the structure of many states in light nuclei, in particular those lying close to the threshold energy for break-up entirely into their constituent $\alpha$ clusters. By considering the nucleus as a mixture of fermions and bosons one may wonder whether the bosonic properties may dominate over the fermionic properties in some instances.  Recently, the idea of  
searching for signatures of boson condensation phenomena in excited nuclear systems has become increasingly important \cite{raduta2011, manfredi2012,bonasera2000, natowitz2010}. Along this direction, it has been suggested that boson condensate signatures may be observed in hot nuclei produced during heavy-ion collisions~\cite{oertzen2010}. By colliding projectile and target nuclei at beam energies of the order of $E/A=15-50$ MeV one can produce excited finite nuclear systems that  decay by break-up into 
one or more complex fragments (Z>2) and light particles ($Z\leqslant 2$). 
The mixture of these nuclear species consisting of  fermions and bosons is still matter of intensive investigations in nuclear physics.
The present article aims at exploring signatures of Bose-Einstein condensation  phenomena in the decay of excited $\alpha$-conjugate or deuteron-conjugate quasi-projectile systems produced in semi-peripheral Ca+Ca collisions at $E/A=35$ MeV. An innovative experimental setup,  combining a 4$\pi$ detector array and a high resolution magnetic spectrometer, has been used to study such excited nuclear systems. The obtained results display analogies with similar phenomena observed when studying atomic traps~\cite{schreck2001}, with interesting links between atomic and nuclear physics phenomenologies.

The experiment was performed at the Grand Accelerateur National d'Ions Lourds (GANIL). $^{40}$Ca targets were bombarded with $^{40}$Ca beams at $35\,$MeV/nucleon  energy. 
The setup was constituted by 288 telescopes of the $4\pi$ detector INDRA \cite{pouthas1995}, covering angles $\theta=7^{o}-176^{o}$, and by the large acceptance and high resolution VAMOS magnetic spectrometer~\cite{savajols2003} at very forward angles ($2^{o}<\theta<7^{o}$), which triggers the data acquisition.
A detailed description of the detectors and their coupling can be found in \cite{mariniIWM09}. The combined setup allowed to reconstruct the mass, charge and excitation energy of the quasi-projectile (QP) system, as well as to characterize its decay channels on an event-by-event basis. Such decay leaves the system with a forward moving QP residue, detected and identified with the VAMOS spectrometer, and coincident light particles and fragments emitted at larger angles and detected by INDRA telescopes. 
Only peripheral and semi-peripheral collisions, leading to a heavy QP remnant detected  in VAMOS with $Z>5$ have been studied in this work. 
In order to reconstruct the charge, $Z_{QP}$, mass, $A_{QP}$, and momentum vector, $   \vec{p}_{QP}$, of the QP, particles with $Z=1$, $2$ and $Z\geq3$, detected by INDRA were attributed to  QP decay when their longitudinal velocities lay within the range of $\pm65\%$, $\pm60\%$, $\pm45\%$, respectively, of the coincident QP residue velocity \cite{steckmeyer01}.
This selection is intended to remove fragments from non-QP sources \cite{mariniNIMA}.
 In order to minimize contributions from collective (entrance-channel) effects, which are predominant in the beam direction, we have estimated the excitation energy, $E^{\star}/A$, of the reconstructed QPs from the momenta of their accompanying emitted particles transverse to the quasi-projectile momentum ($p_{\perp}$). The so-called transverse excitation energy ($E^{\star}$) \cite{wuenschel2010} was calculated through calorimetry as the sum of the charged particle transverse kinetic energy in the QP reference frame ($K_{\perp}^{i}$), corrected for the reaction \textit{Q-}value: $E^{\star} = \frac{3}{2}\sum_{i} K_{\perp}^{i} - Q_{value}$. 
Events were then sorted in 13 excitation energy bins, $0.5\,$MeV/nucleon wide, from 0 to 6.5 MeV/nucleon. 

Events with a reconstructed QP mass between 34 and 46 were selected, which correspond to a QP charge distribution centered at $Z_{QP}=20$ with a standard deviation of about $1$ unit.  
The reconstructed mass A$_{QP}$ does not account for the emitted and not detected neutrons. However, simulations performed with the GEMINI code \cite{charity88_2} show that the evaporation of Ca QPs at these measured excitation energies mostly produces an  average neutron multiplicity $M_n  \lesssim 1$. The uncertainty on $E^{\star}/A$ due to the non-detection of neutrons is within the chosen $E^{\star}/A$ bin width.

\begin{figure}
\centering
\includegraphics[width=0.70\columnwidth]
{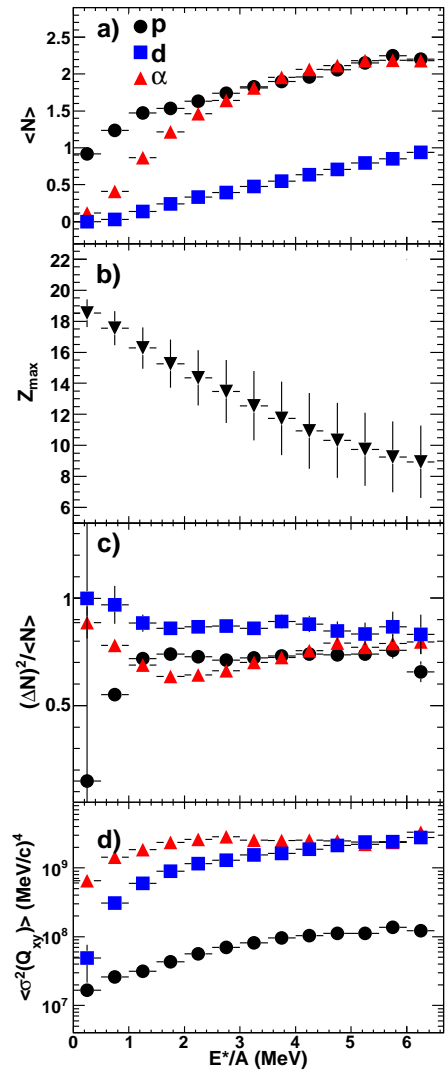}
\caption{(Color online) (a) Measured mean multiplicity of protons, deuterons and alphas; (b)  charge of the largest fragment;  (c) multiplicity fluctuations and (d) quadrupole momentum fluctuations  of particles emitted by QPs as a function of their transverse excitation energy per nucleon, $E^{\star}/A$.}
\label{Fig0_Multiplicities_ZmaxRMS_MultipErrorMean_AndFluct}
\end{figure}

Fig. \ref{Fig0_Multiplicities_ZmaxRMS_MultipErrorMean_AndFluct} (panel a) shows the measured multiplicity of different light particles as a function of the reconstructed transverse excitation energy of the QP. We observe their increase with increasing $E^{\star}$. In panel b we also show the evolution of the charge, $Z_{max}$, of the largest fragment left by the decay of the QP. These results confirm what was already shown in previous works \cite{bonnet2009}: at low excitation energies, the warm QP system decays by evaporating few light particles and leaving a heavy residue, resulting in the observation of larger values for $Z_{max}$; as the excitation energy increases the QP decays by the emission of a larger number of lighter fragments and the maximum observed charge decreases. By placing specific constraints on the quadrupole moment of the emitted particles momentum distribution in the reconstructed QP frame \cite{wuenschel2010,marini2011}, one can isolate events with, on average, isotropic emission. These events are therefore characterized by a certain degree of equilibration. The multiplicity fluctuations for protons (p), deuterons (d) and alphas ($\alpha$), shown in Fig.\ref{Fig0_Multiplicities_ZmaxRMS_MultipErrorMean_AndFluct} (panel c) as a function of the transverse excitation energy per nucleon of the system, are all below the classical limit $\frac{(\Delta N)^{2}}{\langle N \rangle}=1$, obtained when neglecting their quantum nature \cite{landau}. Moreover, significant differences exist between fermions and bosons at low $E^{\star}$. The same is true for the quadrupole momentum ($Q_{xy}$) fluctuations of the three probe particles, shown in Fig. \ref{Fig0_Multiplicities_ZmaxRMS_MultipErrorMean_AndFluct} (panel d), where differences up to about 2 orders of magnitude can be observed between fermions and bosons. The temperatures and local volumes of the produced systems can then be estimated by studying the measured particle quadrupole momentum  and multiplicity fluctuations, as well as mean multiplicities, according to the method described in \cite{zhengNPA892, zhengPLB696}. The fermionic and bosonic nature of the particles, as well as Coulomb repulsion between them, were taken into account \cite{zhengPRC88}. 
In this work we focus on the study of the extracted volumes, while temperatures will be presented in a forthcoming paper \cite{mariniTemperatures}.
In order to select $\alpha$-conjugate and deuteron-conjugate systems, and maximize the probability of observing signals of boson condensation, events with a reconstructed quasi-projectile mass $A_{QP}=40$ were selected. Afterwards,  to increase the statistics, we selected events with $A_{QP}=36,\,40$ and $44$ for the analysis of $\alpha$ particles, with $34\leq A_{QP} \leq 46$ for deuterons, and with 
$38\leq A_{QP}\leq42$ for  protons, with  results similar to those obtained when selecting $A_{QP}=40$ events.

\begin{figure}
\centering
\includegraphics[width=0.8\columnwidth]
{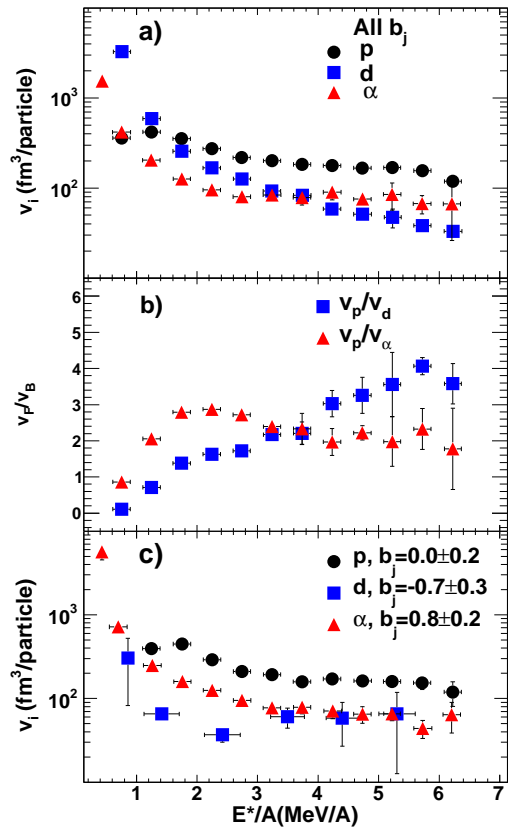}
\caption{(Color online) (a) Mean probed volume per particle ($ \frac{V_{i}}{\langle N \rangle_{i}}$) vs transverse excitation energy per nucleon extracted from proton, deuteron and $\alpha$ fluctuations. (b) Mean probed volume per fermion  normalized to the probed volume per boson. (c) Same as panel (a) with gates applied on $b_j$ values (see text). Panels a and b refer to mixture events without any selection on $b_{j}$.}
\label{Fig1_RhoRho0vsEstar.pdf}
\end{figure}

With the mentioned method, one can extract the local mean volume $V_{i}$ ``as seen'' by a specific probe particle $i$, with $i=$ p, d or $\alpha$.  
By means of the mean multiplicities $\langle N \rangle_{i}$ of  particles measured with the INDRA-VAMOS setup, one can provide simple estimates of the mean partial volumes probed by each of the probe particles, i.e.  $v_{i} = \frac{V_{i}}{\langle N \rangle_{i}}$, which are associated to the local density of particles of species i.
These mean probed volumes,   shown in Fig.  \ref{Fig1_RhoRho0vsEstar.pdf} (panel a) as a function of the excitation energy per nucleon $E^{*}/A$, are much larger than the volume occupied by each particle, $ v_{0,i}$ \cite{commentOnVolumeI}. 
Therefore one can associate them to mean probed volumes in the dilute phase (gas-like) produced by the de-excitation of the hot QP system.

If we  now compare the results shown in Fig. \ref{Fig1_RhoRho0vsEstar.pdf} for bosons (d or $\alpha$) and  fermions (p), we observe that the mean probed volume per protons and alphas decreases as the excitation energy increases, both saturating at $\sim2\,$MeV/nucleon around $200$ and $80$ fm$^{3}$ ($\sim30v_{0,p}$ and $3v_{0,\alpha}$), respectively. 
However, the saturation does not occur for deuterons. Their mean probed volumes are above those extracted with fermions (protons) at low $E^{\star}$, while they drop below the probed volumes per $\alpha$ at higher $E^{\star}$. 
The observed differences between the mean probed volumes for $\alpha$s and d\textit{s} may be associated to different emission mechanisms occurring at different times for the two particle species.
Light particle correlation measurements have indeed shown that emission chronology might exist for different particle species \cite{verde2006}.
Different emission times are not accounted for in the present analysis, which could introduce  space-time ambiguities. Regardless of these details, we observe that the volume probed by fermions is always much larger than the volume  associated to both boson species.
This is more easily observed in panel (b) where the mean probed volume per fermion  normalized to the mean probed volume per bosons, i.e. deuterons and $\alpha$s,   are 
shown as a function of excitation energy. The ratios span values from 1 to 4, depending on $E^{\star}$. 
To better illustrate the consequences of this result, we assume, for simplicity, that the volumes probed by the probe particles have regular shapes (cube, sphere, etc.). In this case the volume would be proportional to the third power of the mean distance R between pairs of a given particle.
Then from the observed difference between volumes we can deduce, on average, estimates of 
mean distance between fermions up to $\sim60\%$ bigger than the mean distance between bosons. 
These behaviours recall what is observed in atomic physics  \cite{schreck2001}: while bosons seem to condense, experiencing  smaller mean volumes and smaller relative distances, fermions, due to the Pauli principle, tend to move apart, experiencing bigger volumes and bigger relative distances.

In the cases on panels a and b of Fig. \ref{Fig1_RhoRho0vsEstar.pdf}, each event is a mixture of interacting fermions and bosons. In order to shed more light into these observations, we try to separate bosonic-like events and fermionic-like events by means of the following event-by-event quantity: 

\begin{equation}\label{eq.bjeq}
b_{j} = \frac{1}{M} \sum_{i=1}^{M}\frac{(-1)^{N_{i}}+(-1)^{Z_{i}} }{2}
\end{equation}
where M is the event multiplicity and $Z_{i}$ and $N_{i}$ are the charge and neutron numbers of the \textit{i}-th fragment, respectively. $b_{j} $ is equal to $1$, $-1$ and $0$  when all fragments emitted in the event are Z even-N even,  Z odd-N odd, and A-odd, respectively.
\begin{figure}
\centering
\includegraphics[width=0.85\columnwidth]{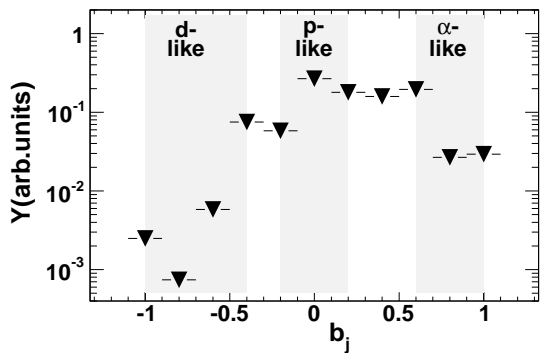} 
\caption{ Distribution of $b_{j}$, as defined in eq.\ref{eq.bjeq}. }
\label{fig:bjDistrib.eps}
\end{figure}
The distribution of this observable obtained for each event is shown on Fig.\ref{fig:bjDistrib.eps}. It allows us to isolate events mostly dominated by the emission of bosons or fermions. We follow the strategy suggested in Ref.~\cite{zhengPRC88}: we select {\it $\alpha$-like, p-like} and {\it d-like} events by applying gates $b_{j}$=$0.8\pm0.2$, $0.0\pm0.2$ and $-0.7\pm0.3$, respectively. Due to lower statistics, we increase the bin size for both $b_{j}$ and $E^{*}/A$ distributions when selecting d-like events.

Fig. \ref{Fig1_RhoRho0vsEstar.pdf} (panel c)  shows the extracted mean volume per particle, $v_{i}$, as a function of excitation energy, $E^{\star}/A$, as seen by protons in p-like events, by deuterons in d-like events  and by alphas in $\alpha$-like events. The mean probed volumes per boson  are in remarkably good agreement for the higher excitation energies and always smaller than the volume per fermions. These results shed more light on the observations presented in  Fig.\ref{Fig1_RhoRho0vsEstar.pdf} panels (a) and (b). Indeed, we observe that the mean probed volume per  $\alpha$ and per proton in $\alpha$-like and p-like events (panel c) are not different from the ones shown on panel (a) and corresponding to events without $b_j$-gating, and thus containing mixtures of bosons and fermions.
 These observations suggest that bosons experience smaller volumes as compared to fermions both in purely boson-like events and in events where mixtures of bosons and fermions are emitted. These  signals, consistent with the possible existence of  boson condensation phenomena,  seem to persist even in the presence of fermions.

In summary we have studied the decay of excited quasi-projectile systems produced in mid-peripheral $^{40}$Ca+$^{40}$Ca collisions at $E/A=35$ MeV with the INDRA-VAMOS setup. Within the selected events, mean partial volumes per particle, probed by bosons (deuterons and alphas) and by fermions (protons) in the low density gas-like phase have been estimated with quantum fluctuation methods. We have selected classes of events where the low density region, originating from the decay of the quasi-projectile, is purely composed by either bosons or fermions and classes of events with mixtures of both types of particles. The observed results seem to show that bosons experience a smaller  volume than fermions. 
These results may be associated to the presence of boson condensation and fermionic quenching phenomena in nuclear systems. These phenomena are observed even in events where mixtures of bosons and fermions coexist, suggesting that these phenomena are not reduced by boson-fermion interactions.
The results of this work recall closely similar phenomenona observed in atomic systems when studying the absorption images of $^{7}$Li (bosons) and $^{6}$Li  (fermions) atomic mixture, measured in a magnetic trap \cite{schreck2001}. In that work the authors observe the coexistence of a quasi-pure Bose Einstein condensate in a Fermi sea.
The similarity of this phenomenon to our results shown on Fig.~\ref{Fig1_RhoRho0vsEstar.pdf} seems to indicate a similar nature for processes occurring in atomic scale and nuclear scale quantum systems, regardless of their different sizes and characteristic interactions, thus stimulating interdisciplinary studies. 
 Future investigations on implications of these phenomena on $\alpha$ clustering and symmetry energy at low density \cite{typel2014, typel2014arxiv} will be further stimulated by attempts to use dynamical techniques, such as particle-particle correlations \cite{verde2006} to estimate emission densities and volumes.

\section{Acknoledgments}
We thank the staff of the GANIL Accelerator facility for their support during the experiment. We also gratefully acknowledge the collaboration of M. Rejmund, A. Navin and F. Farget which made the experiment successful. This work was supported by Le Commissariat \`a l'\'Energie Atomique et aux \'energies alternatives,  Le Centre Nationale de la Recherche Scientifique, Le Minist\`ere de l'\'Education Nationale, and Le Conseil R\'egional de Basse Normandie. 

\bibliographystyle{apsrev}
\bibliography{bibliogr}

\begin{thebibliography}{30}
\bibitem[{com()}]{commentOnVolumeI}
\bibinfo{note}{For example, if we consider the well-known radius-mass number
  relation, $R=1.2A^{1/3}\,$fm and nuclei as being spherical, the volume
  occupied by each particle, $v_{0,i}$, would roughly be of the order of $7$,
  $14$ and $28\,fm^{3}$, respectively, for p, d and $\alpha$ particles. The
  volumes $v_{i}$ available for these particles, as they are shown in panel (a)
  of Fig.\ref{Fig1_RhoRho0vsEstar.pdf}, are much larger than their individual
  volumes, $v_{i}\gg v_{0,i}$.}
\end{thebibliography}

\end{document}